\documentclass[12pt]{article}
\usepackage{graphicx}
\usepackage{amsmath}
\usepackage{amssymb}
\usepackage{caption2}
\begin{document}
\bibliographystyle{prsty}
\begin{center}
{\large {\bf \sc{  Analysis of the vector form factors
$f^+_{K\pi}(Q^2)$  and $f^-_{K\pi}(Q^2)$ with light-cone QCD sum rules  }}} \\[2mm]
Zhi-Gang Wang$^{1}$ \footnote{Corresponding author; E-mail,wangzgyiti@yahoo.com.cn.  }, Shao-Long  Wan$^{2}$     \\
$^{1}$ Department of Physics, North China Electric Power University, Baoding 071003, P. R. China \\
$^{2}$ Department of Modern Physics, University of Science and Technology of China, Hefei 230026, P. R. China \\
\end{center}

\begin{abstract}
In this article, we calculate the vector form factors
$f^+_{K\pi}(Q^2)$ and $f^-_{K\pi}(Q^2)$ within the framework of the
light-cone QCD sum rules approach. The numerical values of the
$f^+_{K\pi}(Q^2)$ are compatible with the existing theoretical
calculations, the central value of the $f^+_{K\pi}(0)$
($f^+_{K\pi}(0)=0.97$) is in excellent agreement with the values
from the chiral perturbation theory and lattice QCD. The values of
the $|f^-_{K\pi}(0)|$ are very large comparing with the theoretical
calculations and experimental data, and can not give any reliable
prediction. At large momentum transfers with $Q^2> 5GeV^2$, the form
factors $f^+_{K\pi}(Q^2)$ and $|f^-_{K\pi}(Q^2)|$ can either take up
the asymptotic behavior of $\frac{1}{Q^2}$ or decrease more quickly
than  $\frac{1}{Q^2}$, more experimental data are needed to select
the ideal sum rules.
\end{abstract}


\section{Introduction}
Semileptonic $K\to\pi\ell \nu$ ($K_{\ell 3}$) decays provide the
most precise determination of the Cabibbo-Kobayashi-Maskawa (CKM)
matrix element $|V_{us}|$ \cite{CKM}. The experimental input
parameters are the semileptonic decay widths and the vector form
factors $f^+_{K\pi}(q^2)$ and $f^-_{K\pi}(q^2)$, which  are
necessary in calculating the phase space integrals. The main
uncertainty in the quantity $|V_{us}f^+_{K\pi}(0)|$ comes from the
unknown shape of the hadronic form factor $f^+_{K\pi}(q^2)$, which
is measurable at $m_l^2<q^2<(m_K-m_\pi)^2$ in the $K_{\ell 3}$
 decays or $(m_K+m_\pi)^2<q^2<m^2_\tau$  in the $\tau \to K\pi\nu$
decays. The experimental data can be fitted to the functions with
either pole models or series expansions, however, systematic errors
are introduced due to the different parameterizations. The
conservation of the vector current implies $f^+_{K\pi}(0)=1$ at zero
momentum transfer \cite{Leutwyler84} , another powerful theoretical
constraint on the $f^+_{K\pi}(0)$ is provided by the $SU(3)$
symmetry of the light pseudoscalar mesons and the Ademollo-Gatto
theorem \cite{Gatto64}, the  $SU(3)$ symmetry  breaking effects
$f^+_{K\pi}(0)-1$ start at  second order in $m_s-m_{q}$. Chiral
perturbation theory (ChPT) provides a natural and powerful tool to
take into account the $SU(3)$ symmetry breaking effects due to the
masses of the light quarks, the $f^+_{K\pi}(q^2)$ is usually
calculated by the ChPT \cite{ChPTKP}. Presently, the comparison
between theory and experiment, and among different experiments, is
complicated by the uncertainties in the form factor
$f^+_{K\pi}(q^2)$ with zero momentum transfer.

In this article, we calculate the values of the vector form factors
 $f^+_{K\pi}(Q^2)$ and $f^-_{K\pi}(Q^2)$ within the framework of the light-cone QCD sum
rules approach. The light-cone QCD sum rules approach carries out
the operator product expansion near the light-cone $x^2\approx 0$
instead of the short distance $x\approx 0$ while the
non-perturbative matrix elements are parameterized by the light-cone
distribution amplitudes
 which classified according to their twists  instead of
 the vacuum condensates \cite{LCSR,LCSRreview}. The non-perturbative
 parameters in the light-cone distribution amplitudes are calculated by   the conventional QCD  sum rules
 and the  values are universal \cite{SVZ79}.

The article is arranged as: in Section 2, we derive the vector  form
factors  $f^+_{K\pi}(Q^2)$ and $f^-_{K\pi}(Q^2)$   with the
light-cone QCD sum rules approach; in Section 3, the numerical
results and discussions; and in Section 4, conclusion.

\section{Vector  form factors  $f^+_{K\pi}(Q^2)$ and $f^-_{K\pi}(Q^2)$   with light-cone QCD sum rules}

In the following, we write down the definitions  for the vector form
factors  $f^+_{K\pi}(q^2)$ and $f^-_{K\pi}(q^2)$,
\begin{eqnarray}
\langle
\pi(q+p)|J_\mu(0)|K(p)\rangle=2f^+_{K\pi}(q^2)p_\mu+\left\{f^+_{K\pi}(q^2)-f^-_{K\pi}(q^2)\right\}q_\mu
\, ,
\end{eqnarray}
where the $J_\mu(x)$ is the vector current. We study the vector form
factors $f^+_{K\pi}(q^2)$ and $f^-_{K\pi}(q^2)$ with the
 two-point correlation functions $\Pi_{\mu}^A(p,q)$ and $\Pi_{\mu}^B(p,q)$,
\begin{eqnarray}
\Pi_{\mu}^A(p,q)&=&i \int d^4x \, e^{-i q \cdot x} \,
\langle 0 |T\left\{J_K(0) J^A_{\mu}(x)\right\}|\pi(p)\rangle \, , \\
\Pi_{\mu}^B(p,q)&=&i \int d^4x \, e^{-i q \cdot x} \,
\langle 0 |T\left\{J_\pi(0) J^B_{\mu}(x)\right\}|K(p)\rangle \, , \\
J^A_\mu(x)&=&{\bar s}(x)\gamma_\mu  u(x)\, ,\nonumber \\
J^{B}_\mu(x)&=&{\bar d}(x)\gamma_\mu  s(x)\, ,\nonumber\\
J_{K}(x)&=& \bar{d}(x)i\gamma_5s(x) \, , \nonumber\\
 J_{\pi}(x)&=& \bar{u}(x)i\gamma_5d(x) \, ,
\end{eqnarray}
 where the $J_K(x)$ and $J_{\pi}(x)$ interpolate    the $K$ and $\pi$ mesons respectively,
 we choose  the pseudoscalar currents to avoid the possible contaminations from the axial-vector mesons.
 The correlation functions
$\Pi^{A(B)}_{\mu}(p,q)$ can be decomposed as
\begin{eqnarray}
\Pi^{A(B)}_{\mu}(p,q)&=&\Pi^{A(B)}_{p}\left(q^2,(q+p)^2\right)p_{\mu}+\Pi^{A(B)}_{q}
\left(q^2,(q+p)^2\right)q_{\mu},
\end{eqnarray}
due to the Lorentz covariance.  In this article, we derive  the sum
rules with the tensor structures $p_\mu$ and $q_\mu$ respectively.

According to the basic assumption of current-hadron duality in the
QCD sum rules approach \cite{SVZ79}, we can insert  a complete
series of intermediate states with the same quantum numbers as the
current operators $J_{K}(x)$  and $J_\pi(x)$  into the correlation
functions $\Pi^A_{\mu} $ and $\Pi^B_{\mu} $ to obtain the hadronic
representation. After isolating the ground state  contributions from
the pole terms of the $K$ and $\pi$   mesons, the correlation
functions $ \Pi^A_\mu$
 and $\Pi^B_\mu$  can be expressed in the following forms,
\begin{eqnarray}
\Pi_{\mu}^A(p,q)&=&\frac{2f_K
m_K^2f^+_{K\pi}(q^2)}{(m_d+m_s)\left\{m_K^2-(q+p)^2\right\}}p_\mu
\nonumber\\
&&+\frac{f_Km_K^2\left\{f^+_{K\pi}(q^2)-f^-_{K\pi}(q^2)\right\}}{(m_d+m_s)\left\{m_K^2-(q+p)^2\right\}}q_\mu+\cdots
\, ,
\\
\Pi_{\mu}^B(p,q)&=&\frac{2f_\pi
m_\pi^2f^+_{K\pi}(q^2)}{(m_d+m_u)\left\{m_\pi^2-(q+p)^2\right\}}p_\mu
\nonumber\\
&& +\frac{f_\pi
m_\pi^2\left\{f^+_{K\pi}(q^2)-f^-_{K\pi}(q^2)\right\}}{(m_d+m_u)\left\{m_\pi^2-(q+p)^2\right\}}q_\mu+\cdots
\, ,
\end{eqnarray}
here we have not shown the contributions from the high resonances
and continuum states explicitly, they are suppressed after  the
Borel transformation and subtraction.
 We use the standard definitions  for the weak  decay constants $f_K$
 and  $f_{\pi}$,
\begin{eqnarray}
\langle0|J_{K}(0)|K(q)\rangle&=& \frac{f_{K}m_K^2}{m_s+m_d} \, ,\nonumber \\
\langle0|J_{\pi}(0)|\pi(q)\rangle&=&
\frac{f_{\pi}m_{\pi}^2}{m_u+m_d}\, .\nonumber
\end{eqnarray}

 In the following, we briefly outline
the operator product expansion for the correlation functions
$\Pi^A_\mu$ and $\Pi^B_\mu$ in perturbative QCD theory. The
calculations are performed at the large space-like momentum regions
$P^2=-(q+p)^2\gg 0$ and  $Q^2=-q^2\gg 0$, which correspond to the
small light-cone distance $x^2\approx 0$   required by the validity
of the operator product expansion approach. We write down the
propagator of a massive quark in the external gluon field in the
Fock-Schwinger gauge firstly \cite{Belyaev94},
\begin{eqnarray}
&&\langle 0 | T \{q_i(x_1)\, \bar{q}_j(x_2)\}| 0 \rangle =
 i \int\frac{d^4k}{(2\pi)^4}e^{-ik(x_1-x_2)}\nonumber\\
 &&\left\{
\frac{\not\!k +m}{k^2-m^2} \delta_{ij} -\int\limits_0^1 dv\, g_s \,
G^{\mu\nu}_{ij}(vx_1+(1-v)x_2)
 \right. \nonumber \\
&&\left. \Big[ \frac12 \frac {\not\!k
+m}{(k^2-m^2)^2}\sigma_{\mu\nu} - \frac1{k^2-m^2}v(x_1-x_2)_\mu
\gamma_\nu \Big]\right\}\, ,
\end{eqnarray}
where the $G_{\mu \nu }$ is the gluonic field strength, the $g_s$
denotes the strong coupling constant. Substituting the above $s$,
$d$ quark propagators and the corresponding $\pi$, $K$ mesons
light-cone distribution amplitudes into the correlation functions
$\Pi^A_\mu$ and $\Pi^B_\mu$ in Eqs.(2-3) and completing the
integrals over the variables $x$ and $k$, finally we obtain the
representations at the level of quark-gluon degrees of freedom,
\begin{eqnarray}
\Pi_p^A&=&  \frac{f_\pi m_\pi^2}{m_u+m_d}\int_0^1du
\frac{u\phi_p(u)}{m_s^2-(q+up)^2}-m_sf_\pi m_\pi^2\int_0^1du
\int_0^u
dt\frac{u B(t)}{\left\{m_s^2-(q+up)^2\right\}^2} \nonumber\\
&&+\frac{1}{6}\frac{f_\pi m_\pi^2}{m_u+m_d}\int_0^1du
\phi_\sigma(u)\left\{\left[1-u\frac{d}{du}
\right]\frac{1}{m_s^2-(q+up)^2}+\frac{2m_s^2}{\left[m_s^2-(q+up)^2\right]^2}\right\}
  \nonumber\\
  &&+m_sf_\pi \int_0^1 du \left\{ \frac{\phi_\pi(u)}{m_s^2-(q+up)^2} -\frac{m_\pi^2m_s^2}{2} \frac{A(u)}{\left[m_s^2-(q+up)^2\right]^3}\right\}\nonumber\\
&&-f_{3\pi}\int_0^1dv \int_0^1d\alpha_g \int_0^{1-\alpha_g}d\alpha_u
T(\alpha_d,\alpha_g,\alpha_u)\nonumber\\
&&\left\{\frac{(1+2v)u m_\pi^2
}{\left[m_s^2-(q+up)^2\right]^2}-2(1-v)\frac{d}{du}\frac{1}{m_s^2-(q+up)^2}\right\}\mid_{u=(1-v)\alpha_g+\alpha_u}
\nonumber\\
&&+4m_sf_\pi m_\pi^4\int_0^1dv v \int_0^1 d\alpha_g\int_0^{\alpha_g}
d\beta\int_0^{1-\beta}d\alpha
\frac{u\Phi(1-\alpha-\beta,\beta,\alpha)}{\left\{m_s^2-(q+up)^2\right\}^3}\mid_{1-v\alpha_g}
\nonumber \\
&& -4m_sf_\pi m_\pi^4\int_0^1 dv\int_0^1
d\alpha_g\int_0^{1-\alpha_g} d\alpha_u
 \int_0^{\alpha_u}d\alpha
\frac{u\Phi(1-\alpha-\alpha_g,\alpha_g,\alpha)}
{\left\{m_s^2-(q+up)^2\right\}^3}\mid_{u=(1-v)\alpha_g+\alpha_u}\nonumber\\
&&+m_s f_\pi m_\pi^2  \int_0^1dv \int_0^1d\alpha_g
\int_0^{1-\alpha_g}d\alpha_u
\frac{\Psi(\alpha_d,\alpha_g,\alpha_u)}{\left\{m_s^2-(q+up)^2\right\}^2}\mid_{u=(1-v)\alpha_g+\alpha_u}
\, ,
\end{eqnarray}

\begin{eqnarray}
\Pi_p^B&=&  \frac{f_K m_K^2}{m_u+m_s}\int_0^1du
\frac{u\phi_p(u)}{m_d^2-(q+up)^2}-m_d f_K m_K^2\int_0^1du \int_0^u
dt\frac{u B(t)}{\left\{m_d^2-(q+up)^2\right\}^2} \nonumber\\
&&+\frac{1}{6}\frac{f_K m_K^2}{m_u+m_s}\int_0^1du
\phi_\sigma(u)\left\{\left[1-u\frac{d}{du}
\right]\frac{1}{m_d^2-(q+up)^2}+\frac{2m_d^2}{\left[m_d^2-(q+up)^2\right]^2}\right\}
  \nonumber\\
  &&+m_df_K \int_0^1 du \left\{ \frac{\phi_K(u)}{m_d^2-(q+up)^2} -\frac{m_K^2m_d^2}{2} \frac{A(u)}{\left[m_d^2-(q+up)^2\right]^3}\right\}\nonumber\\
&&-f_{3K}\int_0^1dv \int_0^1d\alpha_g \int_0^{1-\alpha_g}d\alpha_s
T(\alpha_u,\alpha_g,\alpha_s)\nonumber\\
&&\left\{\frac{(1+2v)u m_K^2
}{\left[m_d^2-(q+up)^2\right]^2}-2(1-v)\frac{d}{du}\frac{1}{m_d^2-(q+up)^2}\right\}\mid_{u=(1-v)\alpha_g+\alpha_s}
\nonumber\\
&&+4m_df_K m_K^4\int_0^1dv v \int_0^1 d\alpha_g\int_0^{\alpha_g}
d\beta\int_0^{1-\beta}d\alpha
\frac{u\Phi(1-\alpha-\beta,\beta,\alpha)}{\left\{m_d^2-(q+up)^2\right\}^3}\mid_{1-v\alpha_g}
\nonumber \\
&& -4m_df_K m_K^4\int_0^1 dv\int_0^1 d\alpha_g\int_0^{1-\alpha_g}
d\alpha_s
 \int_0^{\alpha_s}d\alpha
\frac{u\Phi(1-\alpha-\alpha_g,\alpha_g,\alpha)}
{\left\{m_d^2-(q+up)^2\right\}^3}\mid_{u=(1-v)\alpha_g+\alpha_s}\nonumber\\
&&+m_d f_K m_K^2  \int_0^1dv \int_0^1d\alpha_g
\int_0^{1-\alpha_g}d\alpha_s
\frac{\Psi(\alpha_u,\alpha_g,\alpha_s)}{\left\{m_d^2-(q+up)^2\right\}^2}\mid_{u=(1-v)\alpha_g+\alpha_s}
\, ,
\end{eqnarray}

\begin{eqnarray}
\Pi_q^A&=&  \frac{f_\pi m_\pi^2}{m_u+m_d}\int_0^1du
\frac{\phi_p(u)}{m_s^2-(q+up)^2}-m_sf_\pi m_\pi^2\int_0^1du \int_0^u
dt\frac{B(t)}{\left\{m_s^2-(q+up)^2\right\}^2} \nonumber\\
&&-\frac{1}{6}\frac{f_\pi m_\pi^2}{m_u+m_d}\int_0^1du
\phi_\sigma(u)\frac{d}{du} \frac{1}{m_s^2-(q+up)^2}
  \nonumber\\
  &&-f_{3\pi}m_\pi^2\int_0^1dv \int_0^1d\alpha_g \int_0^{1-\alpha_g}d\alpha_u
T(\alpha_d,\alpha_g,\alpha_u)\frac{1+2v
}{\left\{m_s^2-(q+up)^2\right\}^2}\mid_{u=(1-v)\alpha_g+\alpha_u}
\nonumber\\
&&+4m_sf_\pi m_\pi^4\int_0^1dv v \int_0^1 d\alpha_g\int_0^{\alpha_g}
d\beta\int_0^{1-\beta}d\alpha
\frac{\Phi(1-\alpha-\beta,\beta,\alpha)}{\left\{m_s^2-(q+up)^2\right\}^3}\mid_{1-v\alpha_g}
\nonumber \\
&& -4m_sf_\pi m_\pi^4\int_0^1 dv\int_0^1
d\alpha_g\int_0^{1-\alpha_g} d\alpha_u
 \int_0^{\alpha_u}d\alpha
\frac{\Phi(1-\alpha-\alpha_g,\alpha_g,\alpha)}{\left\{m_s^2-(q+up)^2\right\}^3}\mid_{u=(1-v)\alpha_g+\alpha_u}
 \, ,\nonumber \\
\end{eqnarray}

\begin{eqnarray}
\Pi_q^B&=&  \frac{f_K m_K^2}{m_u+m_s}\int_0^1du
\frac{\phi_p(u)}{m_d^2-(q+up)^2}-m_df_K m_K^2\int_0^1du \int_0^u
dt\frac{B(t)}{\left\{m_d^2-(q+up)^2\right\}^2} \nonumber\\
&&-\frac{1}{6}\frac{f_K m_K^2}{m_u+m_s}\int_0^1du
\phi_\sigma(u)\frac{d}{du} \frac{1}{m_d^2-(q+up)^2}
  \nonumber\\
  &&-f_{3K}m_K^2\int_0^1dv \int_0^1d\alpha_g
  \int_0^{1-\alpha_g}d\alpha_s
T(\alpha_u,\alpha_g,\alpha_s)\frac{1+2v
}{\left\{m_d^2-(q+up)^2\right\}^2}\mid_{u=(1-v)\alpha_g+\alpha_s}
\nonumber\\
&&+4m_df_K m_K^4\int_0^1dv v \int_0^1 d\alpha_g\int_0^{\alpha_g}
d\beta\int_0^{1-\beta}d\alpha
\frac{\Phi(1-\alpha-\beta,\beta,\alpha)}{\left\{m_d^2-(q+up)^2\right\}^3}\mid_{1-v\alpha_g}
\nonumber \\
&& -4m_df_K m_K^4\int_0^1 dv\int_0^1 d\alpha_g\int_0^{1-\alpha_g}
d\alpha_s
 \int_0^{\alpha_s}d\alpha
\frac{\Phi(1-\alpha-\alpha_g,\alpha_g,\alpha)}{\left\{m_d^2-(q+up)^2\right\}^3}\mid_{u=(1-v)\alpha_g+\alpha_s}
\, , \nonumber \\
\end{eqnarray}
where $\Phi=A_\parallel+A_\perp-V_\perp-V_\parallel$ and
$\Psi=2A_\perp-2V_\perp -A_\parallel+V_\parallel$. In calculation,
we have used the two-particle and three-particle $K$ and $\pi$
mesons light-cone distribution amplitudes
\cite{LCSR,LCSRreview,Belyaev94,Ball98,Ball06}, the explicit
expressions of the $K$ meson light-cone distribution amplitudes are
  presented in the appendix, the corresponding ones for
   the $\pi$ meson can be obtained by simple substitution
  of the non-perturbative parameters. The parameters in the
light-cone distribution amplitudes are scale dependent and can be
estimated with the QCD sum rules approach
\cite{LCSR,LCSRreview,Belyaev94,Ball98,Ball06}. In this article, the
energy scale $\mu$ is chosen to be  $\mu=1GeV$.

We take the  Borel transformation with respect to  the variable
$P^2=-(q+p)^2$    for the correlation functions $\Pi_p^{A(B)}$ and
$\Pi_q^{A(B)}$, and obtain the analytical expressions for those
invariant functions. After matching  with the hadronic
representations below the thresholds, we obtain the following four
sum rules for the form factors $f^+_{K\pi}(q^2)$ and
$f^-_{K\pi}(q^2)$,

\begin{eqnarray}
&&\frac{2f_Km_K^2}{m_d+m_s}f^+_{K\pi}(q^2)e^{-\frac{m_K^2}{M^2}} \nonumber\\
&=&  \frac{f_\pi m_\pi^2}{m_u+m_d}\int_{\Delta_A}^1du
\phi_p(u)e^{-DD}-m_sf_\pi m_\pi^2\int_{\Delta_A}^1du \int_0^u
dt \frac{B(t)}{uM^2}e^{-DD}  \nonumber\\
&&+\frac{1}{6}\frac{f_\pi m_\pi^2}{m_u+m_d}\int_{\Delta_A}^1du
\phi_\sigma(u)\left\{\left[1-u\frac{d}{du}
\right]\frac{1}{u}+\frac{2m_s^2}{u^2M^2}\right\}e^{-DD}
  \nonumber\\
  &&+m_sf_\pi \int_{\Delta_A}^1 du \left\{ \frac{\phi_\pi(u)}{u} - \frac{m_\pi^2m_s^2A(u)}{4u^3M^4}\right\}e^{-DD}\nonumber\\
&&-f_{3\pi}\int_0^1dv \int_0^1d\alpha_g \int_0^{1-\alpha_g}d\alpha_u
T(\alpha_d,\alpha_g,\alpha_u)\Theta(u-\Delta_A)\nonumber\\
&&\left\{\frac{(1+2v) m_\pi^2
}{uM^2}-2(1-v)\frac{d}{du}\frac{1}{u}\right\}e^{-DD}\mid_{u=(1-v)\alpha_g+\alpha_u}
\nonumber\\
&&+2m_sf_\pi m_\pi^4\int_0^1dv v \int_0^1 d\alpha_g\int_0^{\alpha_g}
d\beta\int_0^{1-\beta}d\alpha \nonumber\\
&&\frac{\Phi(1-\alpha-\beta,\beta,\alpha)\Theta(u-\Delta_A)}{u^2M^4}e^{-DD}\mid_{u=1-v\alpha_g}
\nonumber \\
&& -2m_sf_\pi m_\pi^4\int_0^1 dv\int_0^1
d\alpha_g\int_0^{1-\alpha_g} d\alpha_u
 \int_0^{\alpha_u}d\alpha \nonumber\\
 &&\frac{\Phi(1-\alpha-\alpha_g,\alpha_g,\alpha)\Theta(u-\Delta_A)}
{u^2M^4}e^{-DD}\mid_{u=(1-v)\alpha_g+\alpha_u}\nonumber\\
&&+m_s f_\pi m_\pi^2  \int_0^1dv \int_0^1d\alpha_g
\int_0^{1-\alpha_g}d\alpha_u \nonumber\\
&&\frac{\Psi(\alpha_d,\alpha_g,\alpha_u)\Theta(u-\Delta_A)}{u^2M^2}e^{-DD}\mid_{u=(1-v)\alpha_g+\alpha_u}
\, ,
\end{eqnarray}

\begin{eqnarray}
&&\frac{2f_\pi
m_\pi^2}{m_u+m_d}f^+_{K\pi}(q^2)e^{-\frac{m_\pi^2}{M^2}}\nonumber\\
 &=&  \frac{f_K m_K^2}{m_u+m_s}\int_{\Delta_B}^1du
\phi_p(u)e^{-EE}-m_df_K m_K^2\int_{\Delta_B}^1du \int_0^u
dt \frac{B(t)}{uM^2}e^{-EE}  \nonumber\\
&&+\frac{1}{6}\frac{f_K m_K^2}{m_u+m_s}\int_{\Delta_B}^1du
\phi_\sigma(u)\left\{\left[1-u\frac{d}{du}
\right]\frac{1}{u}+\frac{2m_d^2}{u^2M^2}\right\}e^{-EE}
  \nonumber\\
  &&+m_df_K \int_{\Delta_B}^1 du \left\{ \frac{\phi_K(u)}{u} - \frac{m_K^2m_d^2A(u)}{4u^3M^4}\right\}e^{-EE}\nonumber\\
&&-f_{3K}\int_0^1dv \int_0^1d\alpha_g \int_0^{1-\alpha_g}d\alpha_s
T(\alpha_u,\alpha_g,\alpha_s)\Theta(u-\Delta_B)\nonumber\\
&&\left\{\frac{(1+2v) m_K^2
}{uM^2}-2(1-v)\frac{d}{du}\frac{1}{u}\right\}e^{-EE}\mid_{u=(1-v)\alpha_g+\alpha_s}
\nonumber\\
&&+2m_df_K m_K^4\int_0^1dv v \int_0^1 d\alpha_g\int_0^{\alpha_g}
d\beta\int_0^{1-\beta}d\alpha \nonumber\\
&&\frac{\Phi(1-\alpha-\beta,\beta,\alpha)\Theta(u-\Delta_B)}{u^2M^4}e^{-EE}\mid_{u=1-v\alpha_g}
\nonumber \\
&& -2m_df_K m_K^4\int_0^1 dv\int_0^1 d\alpha_g\int_0^{1-\alpha_g}
d\alpha_s
 \int_0^{\alpha_s}d\alpha \nonumber\\
 &&\frac{\Phi(1-\alpha-\alpha_g,\alpha_g,\alpha)\Theta(u-\Delta_B)}
{u^2M^4}e^{-EE}\mid_{u=(1-v)\alpha_g+\alpha_s}\nonumber\\
&&+m_d f_K m_K^2  \int_0^1dv \int_0^1d\alpha_g
\int_0^{1-\alpha_g}d\alpha_s \nonumber\\
&&\frac{\Psi(\alpha_u,\alpha_g,\alpha_s)\Theta(u-\Delta_B)}{u^2M^2}e^{-EE}\mid_{u=(1-v)\alpha_g+\alpha_s}
\, ,
\end{eqnarray}

\begin{eqnarray}
&&\frac{f_K
m_K^2}{m_d+m_s}\left\{f^+_{K\pi}(q^2)-f^-_{K\pi}(q^2)\right\}e^{-\frac{m_K^2}{M^2}}
\nonumber\\
  &=&  \frac{f_\pi
m_\pi^2}{m_u+m_d}\int_{\Delta_A}^1du
\frac{\phi_p(u)}{u}e^{-DD}-m_sf_\pi m_\pi^2\int_{\Delta_A}^1du
\int_0^u
dt\frac{B(t)}{u^2M^2}e^{-DD} \nonumber\\
&&-\frac{1}{6}\frac{f_\pi m_\pi^2}{m_u+m_d}\int_{\Delta_A}^1du
\phi_\sigma(u)\frac{d}{du} \frac{1}{u}e^{-DD}
  \nonumber\\
  &&-f_{3\pi}m_\pi^2\int_0^1dv \int_0^1d\alpha_g \int_0^{1-\alpha_g}d\alpha_u
T(\alpha_d,\alpha_g,\alpha_u)\nonumber\\
&&\Theta(u-\Delta_A)\frac{1+2v
}{u^2M^2}e^{-DD}\mid_{u=(1-v)\alpha_g+\alpha_u}
\nonumber\\
&&+2m_sf_\pi m_\pi^4\int_0^1dv v \int_0^1 d\alpha_g\int_0^{\alpha_g}
d\beta\int_0^{1-\beta}d\alpha \nonumber \\
&& \frac{\Phi(1-\alpha-\beta,\beta,\alpha)
\Theta(u-\Delta_A)}{u^3M^4}e^{-DD}\mid_{1-v\alpha_g}
\nonumber \\
&& -2m_sf_\pi m_\pi^4\int_0^1 dv\int_0^1
d\alpha_g\int_0^{1-\alpha_g} d\alpha_u
 \int_0^{\alpha_u}d\alpha \nonumber\\
 &&\frac{\Phi(1-\alpha-\alpha_g,\alpha_g,\alpha)\Theta(u-\Delta_A)}{u^3M^4}e^{-DD}\mid_{u=(1-v)\alpha_g+\alpha_u}
\, ,
\end{eqnarray}

\begin{eqnarray}
&&\frac{f_\pi
m_\pi^2}{m_u+m_d}\left\{f^+_{K\pi}(q^2)-f^-_{K\pi}(q^2)\right\}e^{-\frac{m_\pi^2}{M^2}}
\nonumber \\
 &=&  \frac{f_K m_K^2}{m_u+m_s}\int_{\Delta_B}^1du
\frac{\phi_p(u)}{u}e^{-EE}-m_df_K m_K^2\int_{\Delta_B}^1du \int_0^u
dt\frac{B(t)}{u^2M^2}e^{-EE} \nonumber\\
&&-\frac{1}{6}\frac{f_K m_K^2}{m_u+m_s}\int_{\Delta_B}^1du
\phi_\sigma(u)\frac{d}{du} \frac{1}{u}e^{-EE}
  \nonumber\\
  &&-f_{3K}m_K^2\int_0^1dv \int_0^1d\alpha_g
  \int_0^{1-\alpha_g}d\alpha_s
T(\alpha_u,\alpha_g,\alpha_s)\nonumber\\
&&\Theta(u-\Delta_B)\frac{1+2v
}{u^2M^2}e^{-EE}\mid_{u=(1-v)\alpha_g+\alpha_s}
\nonumber\\
&&+2m_df_K m_K^4\int_0^1dv v \int_0^1 d\alpha_g\int_0^{\alpha_g}
d\beta\int_0^{1-\beta}d\alpha\nonumber\\
&&\frac{\Phi(1-\alpha-\beta,\beta,\alpha)\Theta(u-\Delta_B)}{u^3M^4}e^{-EE}\mid_{1-v\alpha_g}
\nonumber \\
&& -2m_df_K m_K^4\int_0^1 dv\int_0^1 d\alpha_g\int_0^{1-\alpha_g}
d\alpha_s
 \int_0^{\alpha_s}d\alpha \nonumber\\
 &&\frac{\Phi(1-\alpha-\alpha_g,\alpha_g,\alpha)\Theta(u-\Delta_B)}{u^3M^4}e^{-EE}\mid_{u=(1-v)\alpha_g+\alpha_s}
\, ,
\end{eqnarray}
where
\begin{eqnarray}
DD &=& \frac{m_s^2+u(1-u)m_\pi^2-(1-u)q^2}{uM^2} \, , \nonumber\\
EE &=& \frac{m_d^2+u(1-u)m_K^2-(1-u)q^2}{uM^2} \, , \nonumber\\
\Delta_A&=&\frac{m_s^2-q^2}{s_K^0-q^2} \, , \nonumber\\
\Delta_B&=&\frac{m_d^2-q^2}{s_\pi^0-q^2}  \, ,
\end{eqnarray}
 here   the  $s^0_K$ and $s^0_\pi$ are threshold parameters for the interpolating currents
$J_{K}(x)$  and $J_{\pi}(x)$  respectively.

\section{Numerical results and discussions}
The input parameters of the light-cone distribution amplitudes are
taken as $\lambda_3=1.6\pm0.4$, $f_{3K}=(0.45\pm0.15)\times
10^{-2}GeV^2$, $\omega_3=-1.2\pm0.7$, $\omega_4=0.2\pm0.1$,
$a_2=0.25\pm 0.15$, $a_1=0.06\pm 0.03$, $\eta_4=0.6\pm0.2$  for the
$K$ meson;   $\lambda_3=0.0$, $f_{3\pi}=(0.45\pm0.15)\times
10^{-2}GeV^2$, $\omega_3=-1.5\pm0.7$, $\omega_4=0.2\pm0.1$,
$a_2=0.25\pm 0.15$, $a_1=0.0 $, $\eta_4=10.0\pm3.0$   for the $\pi$
meson \cite{LCSR,LCSRreview,Belyaev94,Ball98,Ball06}; and
$m_s=(137\pm 27 )MeV$, $m_u=m_d=(5.6\pm 1.6)  MeV$, $f_K=0.160GeV$,
$f_\pi=0.130GeV$, $m_K=498MeV$, $m_{\pi} =135MeV$. The threshold
parameters are chosen to be $s^0_{K}=1.1GeV^2$ and
$s^0_{\pi}=0.8GeV^2$, which can reproduce the values of the decay
constants $f_K=160MeV$ and $f_{\pi}=130MeV$ in the QCD sum rules.

The Borel parameters in the four sum rules (see Eqs.(13-16)) are
taken as $M^2=(1-2) GeV^2$, in this region, the values of the form
factors $f^+_{K\pi}(Q^2)$ from Eqs.(13-14) and the
$f^+_{K\pi}(Q^2)-f^-_{K\pi}(Q^2)$ from Eq.(15) are rather stable,
the values of the $f^+_{K\pi}(Q^2)-f^-_{K\pi}(Q^2)$ from Eq.(16) are
not as stable as the ones from Eqs.(13-15), which are shown, for
example, in Fig.1 and Fig.2 respectively. In this article, we take
the special value $M^2=1.5GeV^2$ in   numerical calculations,
although such a definite Borel parameter can not take into account
some uncertainties, the predictive power can not be impaired
qualitatively.
\begin{figure}
\centering
  \includegraphics[totalheight=7cm,width=7cm]{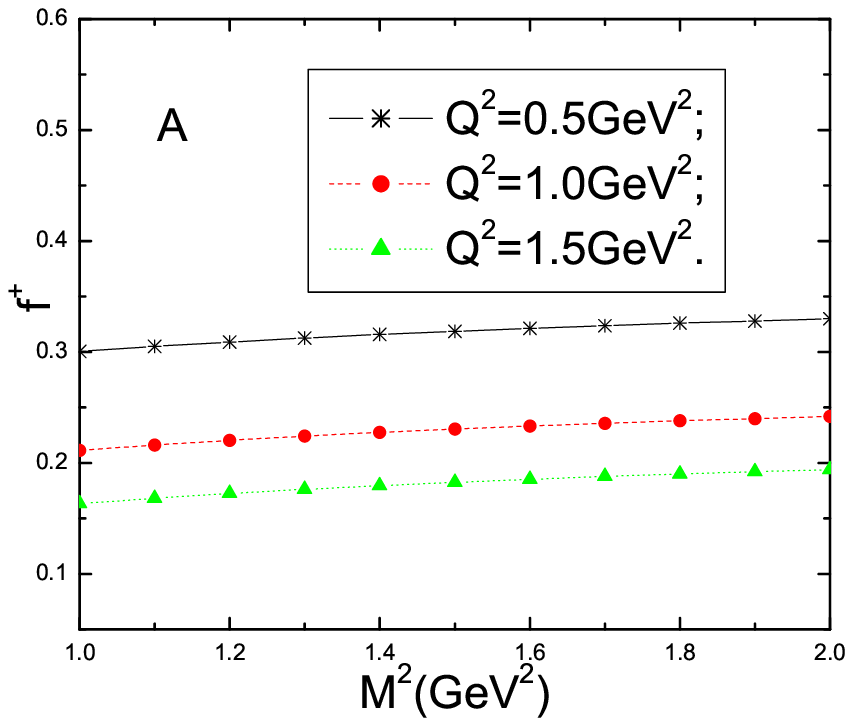}
  \includegraphics[totalheight=7cm,width=7cm]{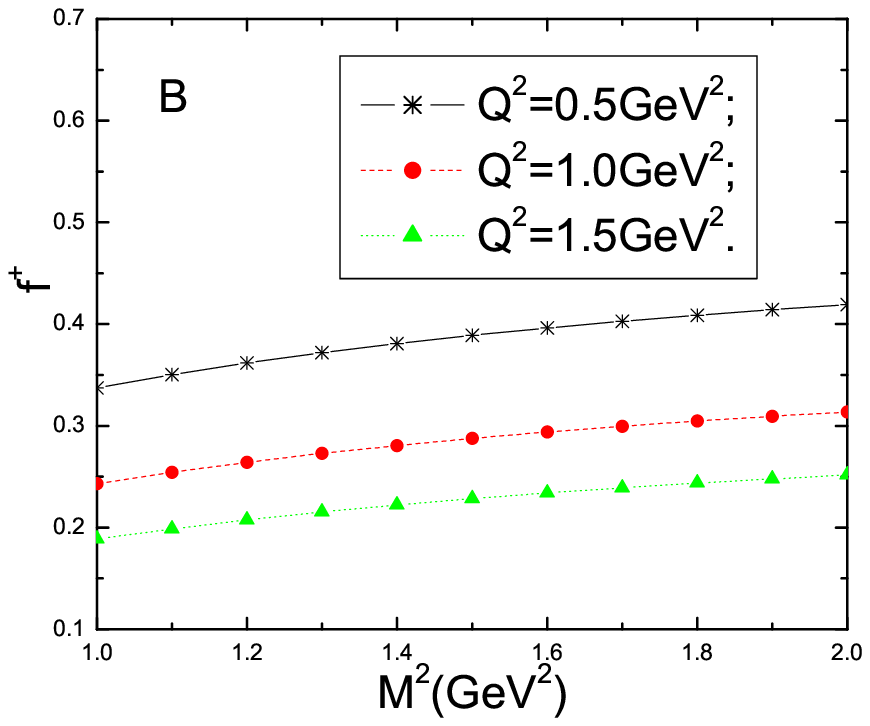}
   \caption{The   $f^+_{K\pi}(Q^2)$ with the parameter $M^2$,
 A from Eq.(13) and B from Eq.(14). }
\end{figure}

\begin{figure}
\centering
  \includegraphics[totalheight=7cm,width=7cm]{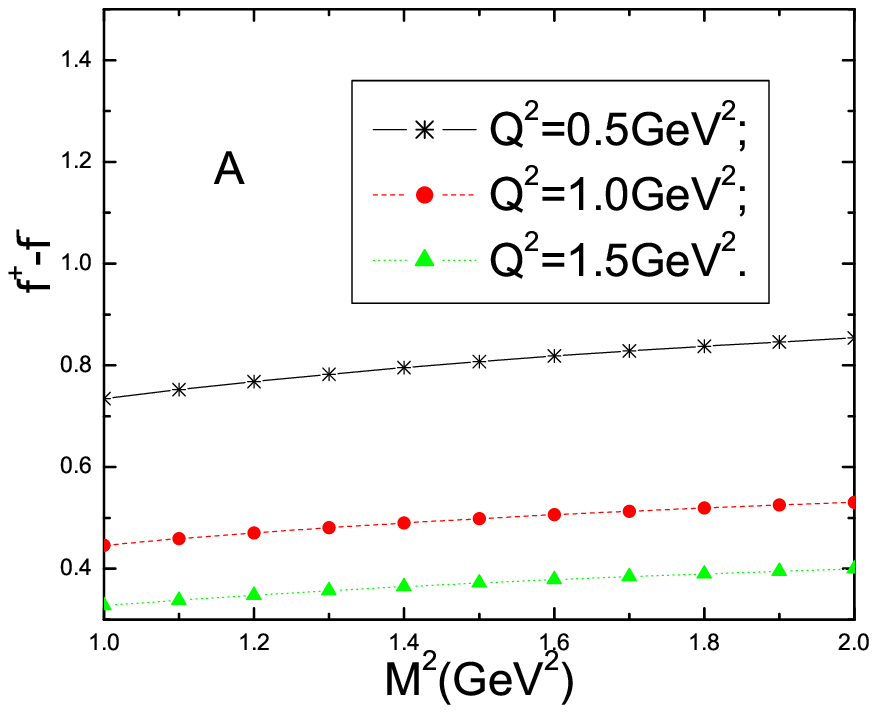}
  \includegraphics[totalheight=7cm,width=7cm]{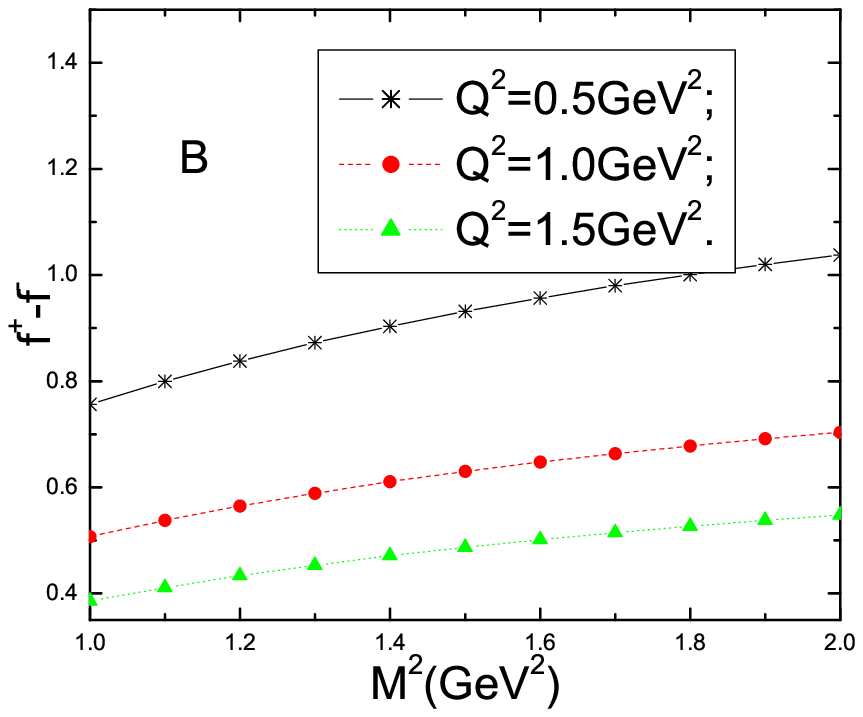}
   \caption{The   $f^+_{K\pi}(Q^2)-f^-_{K\pi}(Q^2)$ with the parameter $M^2$,
 A from Eq.(15) and  B from Eq.(16).  }
\end{figure}

The uncertainties of the seven parameters $f_{3K}$($f_{3\pi}$),
$a_2$, $a_1$, $\lambda_3$, $\omega_3$, $\omega_4$ and $\eta_4$  can
only result in small uncertainties for the numerical values. The
main uncertainties come from the two parameters $m_s$ and
$m_q(=m_u=m_d)$, the variations of those parameters can lead to
large changes for the numerical values, which are shown, for
example, in Fig.3 and Fig.4, respectively.

\begin{figure}
\centering
  \includegraphics[totalheight=7cm,width=7cm]{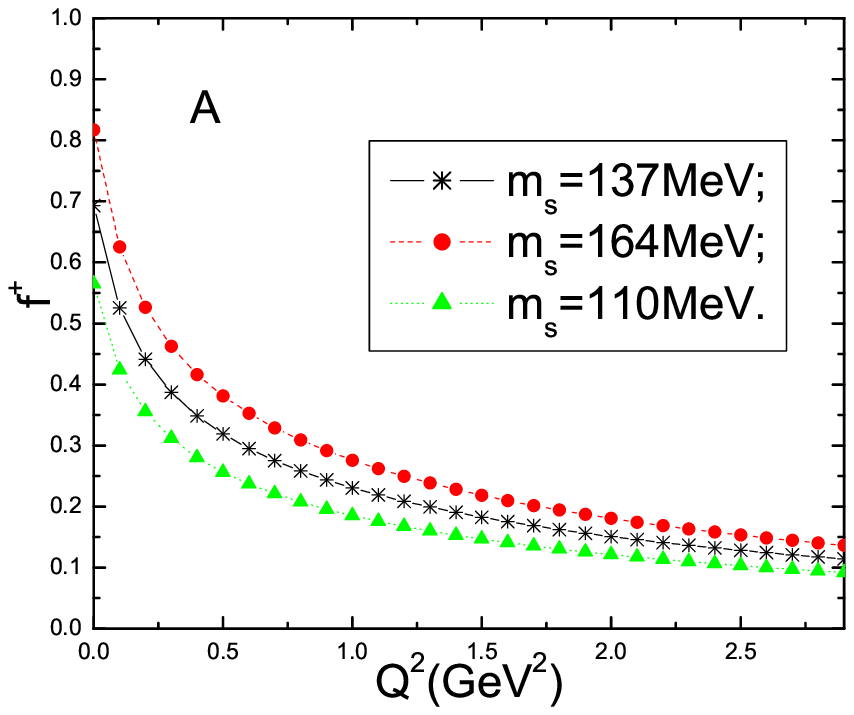}
  \includegraphics[totalheight=7cm,width=7cm]{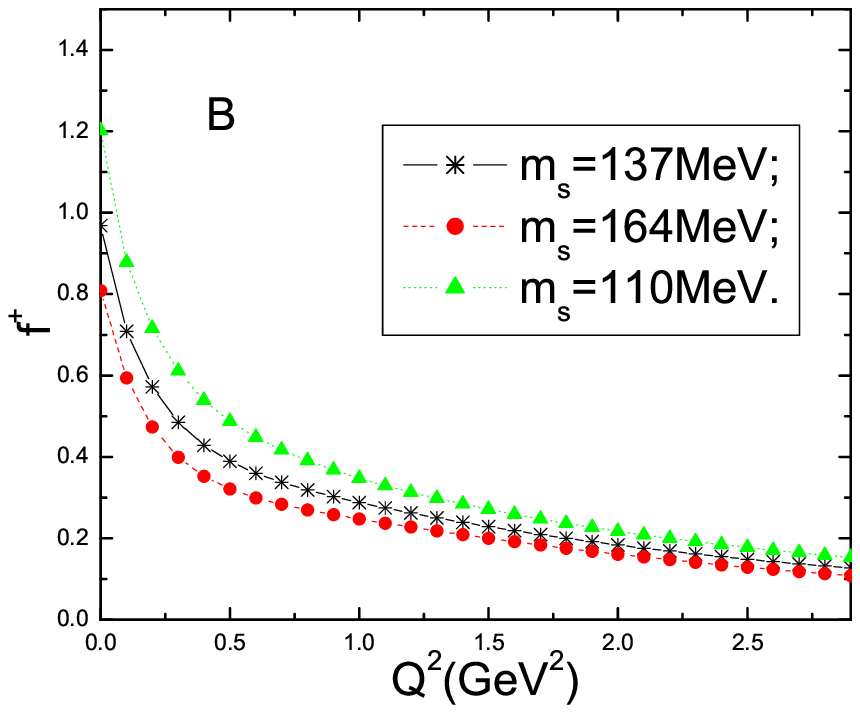}
   \caption{The $f^+_{K\pi}(Q^2)$ with the parameter $m_s$,
 A from Eq.(13) and B from Eq.(14). }
\end{figure}

\begin{figure}
\centering
  \includegraphics[totalheight=7cm,width=7cm]{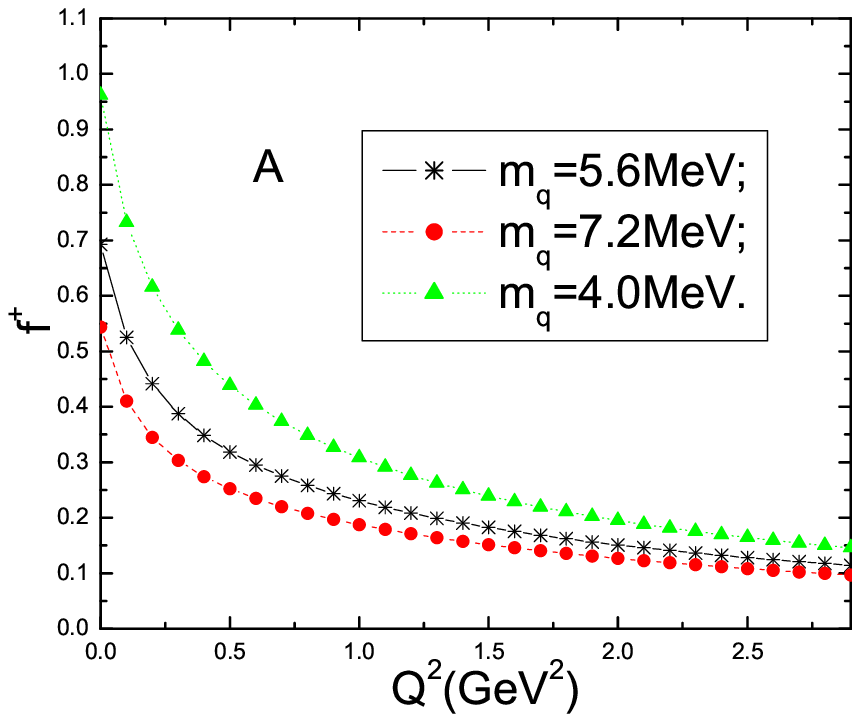}
  \includegraphics[totalheight=7cm,width=7cm]{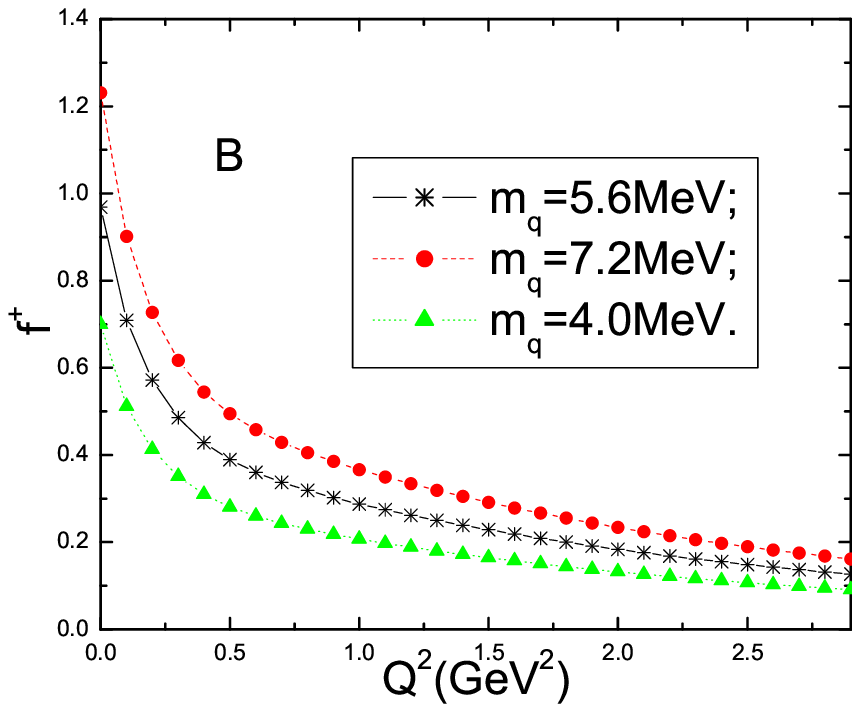}
   \caption{The $f^+_{K\pi}(Q^2)$ with the parameter $m_q$,
 A from Eq.(13) and B from Eq.(14).  }
\end{figure}

From the two sum rules in Eqs.(13-14), we can see that due to the
pseudoscalar currents we choose  to interpolate the $K$ and $\pi$
mesons,  the main contributions come from the two-particle twist-3
light-cone distribution amplitudes,  not  the twist-2 light-cone
distribution amplitudes, those channels can be used to evaluate the
non-perturbative  parameters in the twist-3 light-cone distribution
amplitudes with the experimental data.  The dominating contributions
to the nucleons light-cone distribution amplitudes come from the
three valence quarks, additional contributions from the gluons and
quark-antiquark pairs are very small \cite{DFJK}, the main
contributions to the pseudoscalar mesons light-cone distribution
amplitudes come from the two valence quarks, the two cases are
analogous. In the light-cone QCD sum rules, we can neglect the
contributions from the light-cone distribution amplitudes with
additional valence gluon (or quark-antiquark pair) and make
relatively rough estimations.  For the heavy-light form factors $B
\to \pi, K$, we use the pseudoscalar current  to interpolate the $B$
meson in the framework of the light-cone QCD sum rules, the current
mass of the $b$ quark is very large, we can take the chiral limit
for the masses of the $K$ and $\pi$ mesons \cite{BKaonPion}, the
contributions from the two-particle twist-3 light-cone distribution
amplitudes are very small and can be safely neglected, the
analytical expressions are
 simple.  If we use the axial-vector currents to interpolate the
$K$ and $\pi$ mesons, the tensor structures are more complex, some
structures will get dominating contributions from the twist-2
light-cone distribution amplitudes \footnote{The results with the
axial-vector currents will be presented elsewhere.}.

We  obtain the values of the $f^+_{K\pi}(Q^2)$ from the two sum
rules in Eqs.(13-14),  then take  those values as input parameters,
we can  obtain the  $f^-_{K\pi}(Q^2)$ from the two sum rules  in
Eqs.(15-16).

Taking into account all the uncertainties, finally we obtain the
numerical values of the form factors $f^+_{K\pi}(Q^2)$ and
$f^-_{K\pi}(Q^2)$, which are shown in the Fig.5, at  zero momentum
transfer,
\begin{eqnarray}
 f^+_{K\pi}(0) =0.69^{+0.30}_{-0.20}    \, , \nonumber \\
 f^+_{K\pi}(0) =0.97^{+0.35}_{-0.31}  \, ,  \nonumber \\
  |f^-_{K\pi}(0)| =5.15^{+1.59}_{-0.91}  \, ,  \nonumber \\
|f^-_{K\pi}(0)| =11.47^{+4.98}_{-4.44} \, ,
\end{eqnarray}
from Eq.(13), Eq.(14), Eq.(15) and Eq.(16) respectively.  From the
Fig.5, we can see that the uncertainties are rather large, we should
refine the input parameters $m_s$ and $m_q$ especially the $m_q$ to
improve the predictive ability.
\begin{figure}
\centering
  \includegraphics[totalheight=7cm,width=7cm]{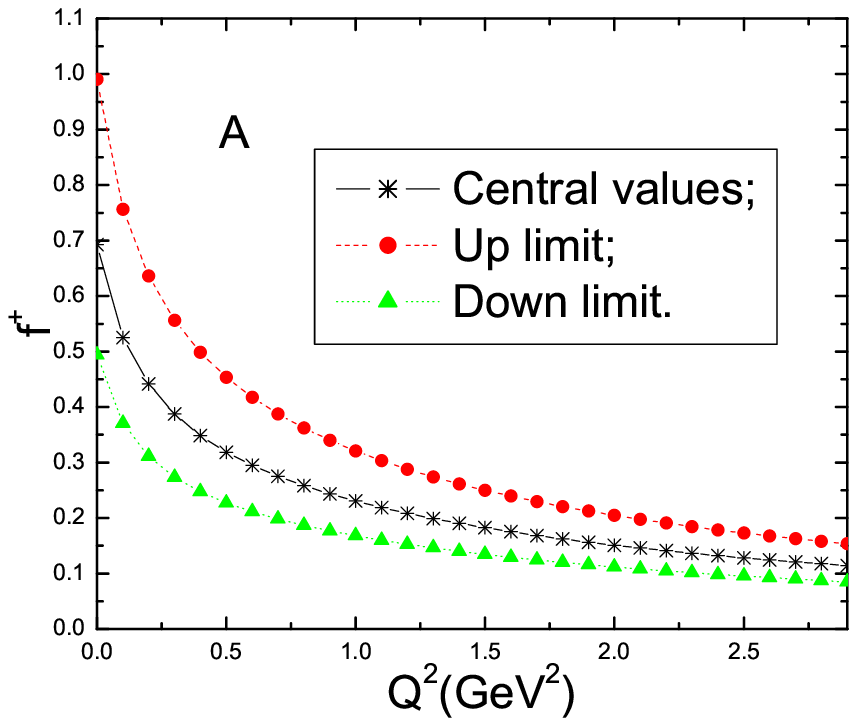}
  \includegraphics[totalheight=7cm,width=7cm]{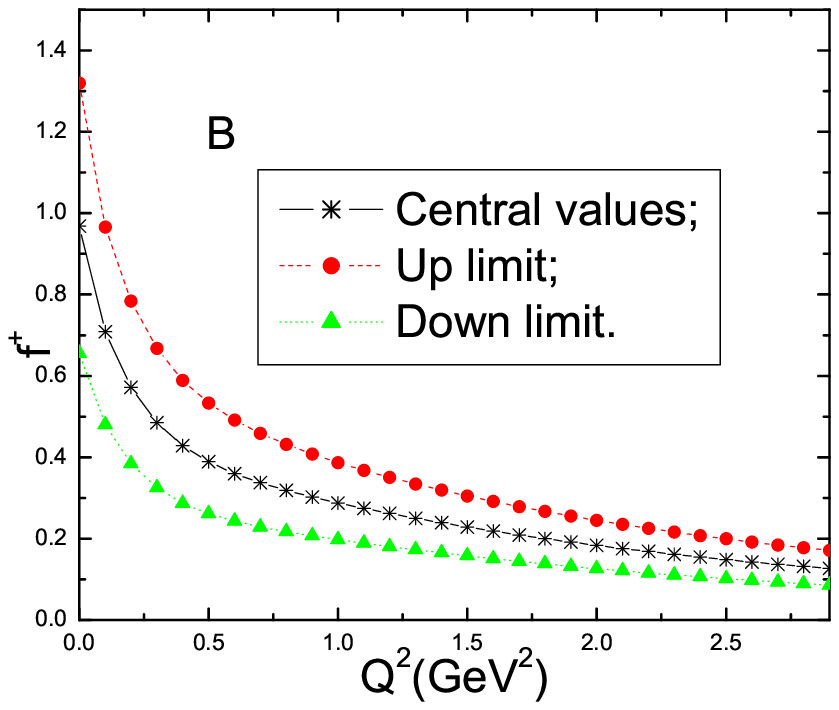}
  \includegraphics[totalheight=7cm,width=7cm]{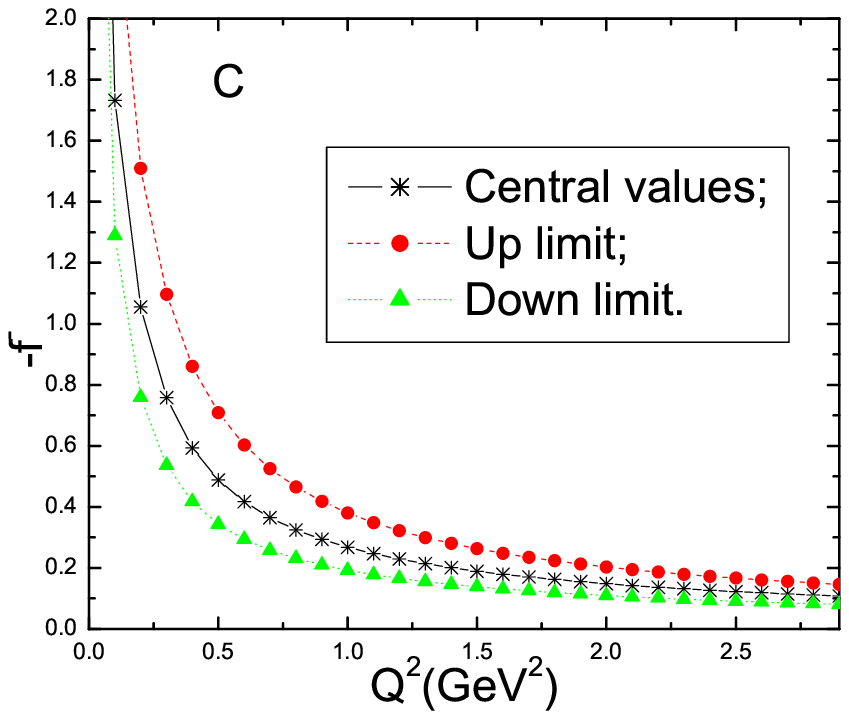}
  \includegraphics[totalheight=7cm,width=7cm]{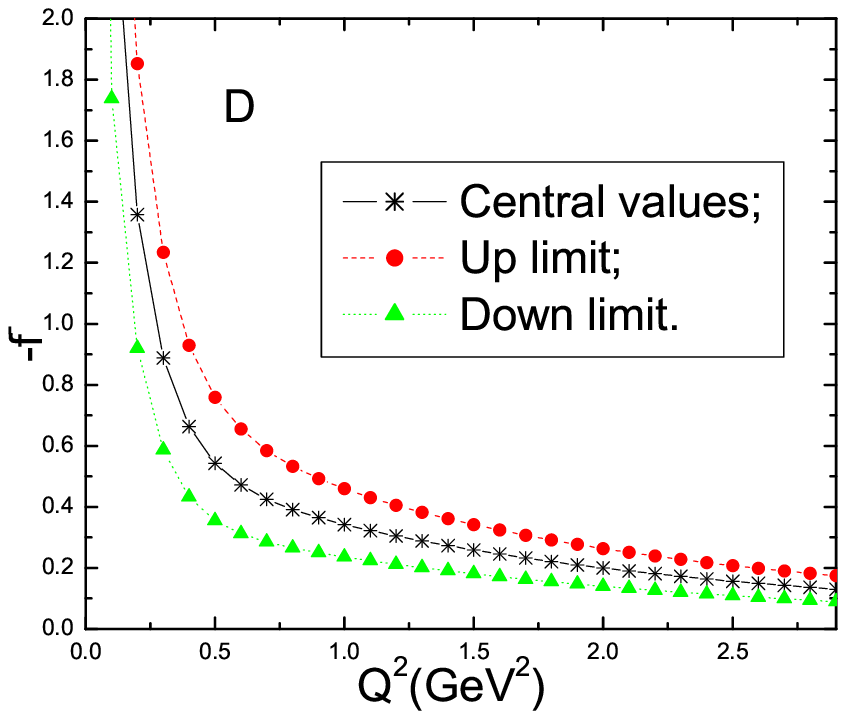}
   \caption{The values of the $f^+_{K\pi}(Q^2)$ and $f^-_{K\pi}(Q^2)$,
 A from Eq.(13), B from Eq.(14), C from Eq.(15) and Eq.(13) , D from Eq.(16) and Eq.(14). }
\end{figure}

The form factors $f^+_{K\pi}(q^2)$ and $f^0_{K\pi}(q^2)$
\footnote{$f^0_{K\pi}(q^2)=f^+_{K\pi}(q^2)+\frac{q^2}{m_K^2-m_\pi^2}f^-_{K\pi}(q^2)$,
the current algebra predicts the value of the scalar form factor
$f^0_{K\pi}(\Delta)$ be $f^0_{K\pi}(\Delta)= -f_K/f_\pi$ at the
Callan-Treiman point $\Delta=m_K^2-m_\pi^2$ \cite{CT66}. } are
measured in the $K_{\ell 3}$ decays with the squared momentum
$q^2>m_l^2$ transfer to the leptons. The curves (or shapes) of the
form factors are always parameterized by the linear model, quadratic
model and pole models to carry out the integrals in the phase space,
the normalization is always chosen to be $f^+_{K\pi}(0)$, i.e.
$f^+_{K\pi}(q^2)=f^+_{K\pi}(0)\left\{1+\lambda_1q^2+\lambda_2q^4+\cdots\right\}$,
etc, the parameters $\lambda_1$, $\lambda_2$, $\cdots$ can be fitted
by the $\chi^2$, etc \cite{ExpKP}. From the experimental data, we
can obtain the values of the $f^+_{K\pi}(0)|V_{us}|$, the basic
parameter $f^+_{K\pi}(0)$ has  to be  calculated with some
theoretical approaches to extract the CKM matrix element $|V_{us}|$.

Comparing with the theoretical calculations from the ChPT
\cite{ChPTKP}  and lattice QCD \cite{LattKP}, the central value of
the  $f^+_{K\pi}(0)$ ($f^+_{K\pi}(0)=0.97$) from Eq.(14)    is
excellent while the value $f^+_{K\pi}(0)=0.70$ from Eq.(13) is
somewhat smaller. The vector form factor $f^+_{K\pi}(Q^2)$ has been
calculated by the ChPT \cite{ChPTKP}, lattice QCD \cite{LattKP}, QCD
sum rules \cite{LCQCDSRKP}, the Bethe-Salpeter equation
\cite{BSEKP},
  etc. The numerical values $f^+_{K\pi}(0)=0.97^{+0.35}_{-0.31}$
from Eq.(14) are more reasonable than the ones from Eq.(13).

In  Figs.6-7, we plot the form factors $f^+_{K\pi}(Q^2)$ and
$f^-_{K\pi}(Q^2)$ at the momentum range $Q^2=(0-12)GeV^2$, from the
figures, we can see that the curve (or shape) of the
$Q^2f^+_{K\pi}(Q^2)$ from Eq.(13) is rather flat at $Q^2> 5 GeV^2 $,
which means that at large momentum transfers, the $f^+_{K\pi}(Q^2)$
takes the asymptotic behavior $f^+_{K\pi}(Q^2)\sim \frac{1}{Q^2}$,
it is expected from the naive power counting rules \cite{Brodsky},
the terms proportional to $\frac{1}{Q^{2n}}$ with $n\geq 2$ are
canceled out with each other.
 The scalar form factor, axial form factor and induced pseudoscalar
form factor of the nucleons take  up the behavior
 $\frac{1}{Q^4}$ at large $Q^2$ \cite{WWY}, which are also expected
 from the naive power counting rules \cite{Brodsky}.
 The curve (or shape) of the $Q^2f^+_{K\pi}(Q^2)$ from Eq.(14) at $Q^2<5GeV^2$ is analogous  to
the electromagnetic form factors of the $K$ and $\pi$ mesons
\cite{Exp06PhD} \footnote{One can consult the thesis \cite{Exp06PhD}
for more literatures on the present states of experimentally
determined electromagnetic form factors of the $\pi$, $K$ and the
proton.}. Because of the $SU(3)$ symmetry of the light flavor
quarks, we expect the vector form factor $f^+_{K\pi}(Q^2)$  will not
have much difference from  the electromagnetic form factors of the
$K$ and $\pi$ mesons \cite{QCDSRPion,SDEPion}, the results from
Eq.(14) at low $Q^2$ are more reasonable than the ones from Eq.(13).
The electromagnetic form factors of the $K$ and $\pi$ mesons have
been calculated with the Bethe-Salpeter equation \cite{SDEPion},
ChPT \cite{ChPTKaonPion},  QCD sum rules\cite{LCQCDSRKP,QCDSRPion},
perturbative QCD \cite{pQCDKaonPion,HnLi},  etc, our numerical
values are compatible  with those theoretical calculations. At large
momentum transfers with $Q^2>5GeV^2$, the terms of the
$f^+_{K\pi}(Q^2)$  proportional to $\frac{1}{Q^{2n}}$ with $n\geq 2$
from Eq.(14) manifest themselves, which result in the curve (or
shape) of the $Q^2f^+_{K\pi}(Q^2)$ decreases  with the increase of
the $Q^2$. The curve (or shape) of the form factor $f^+_{K\pi}(Q^2)$
from Eq.(14) decreases more quickly than  the one from Eq.(13) with
the increase of the $Q^2$.

\begin{figure}
\centering
  \includegraphics[totalheight=7cm,width=7cm]{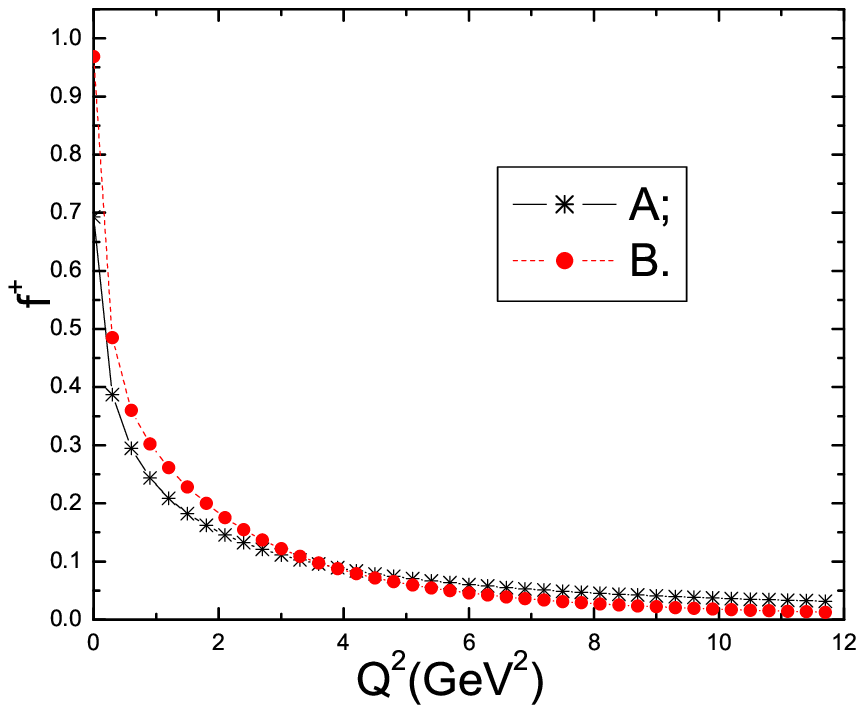}
  \includegraphics[totalheight=7cm,width=7cm]{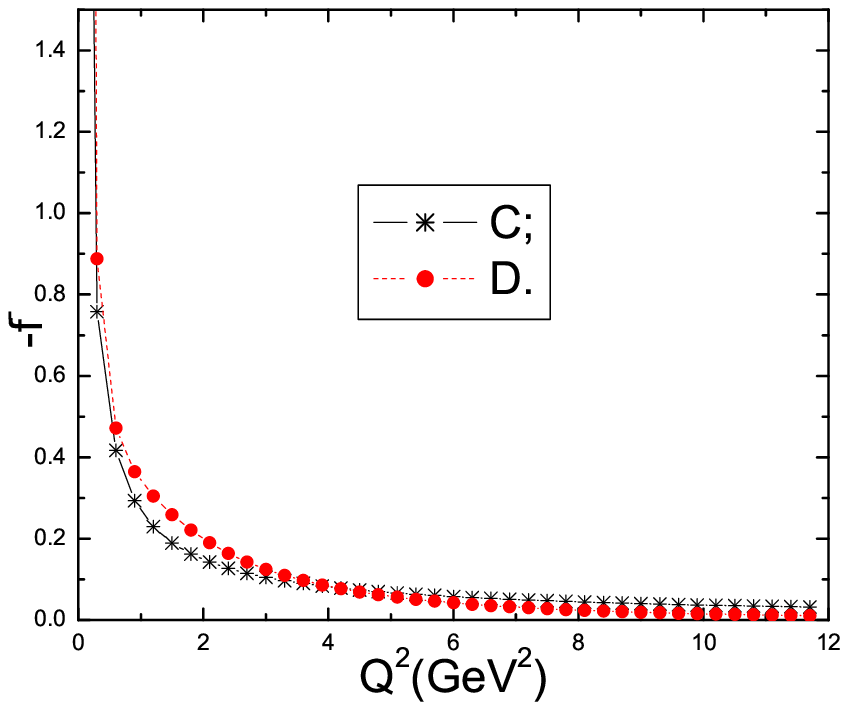}
   \caption{The central values of the $f^+_{K\pi}(Q^2)$ and $f^-_{K\pi}(Q^2)$ at $Q^2=(0-12)GeV^2$,
 A from Eq.(13), B from Eq.(14), C from Eq.(15) and Eq.(13) , D from Eq.(16) and Eq.(14). }
\end{figure}

\begin{figure}
\centering
  \includegraphics[totalheight=7cm,width=7cm]{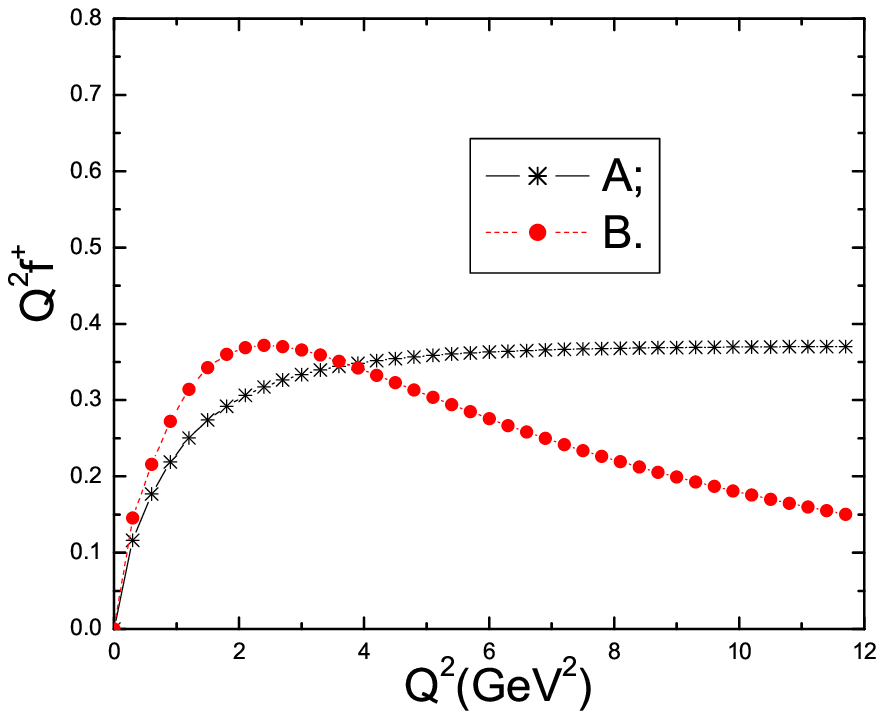}
  \includegraphics[totalheight=7cm,width=7cm]{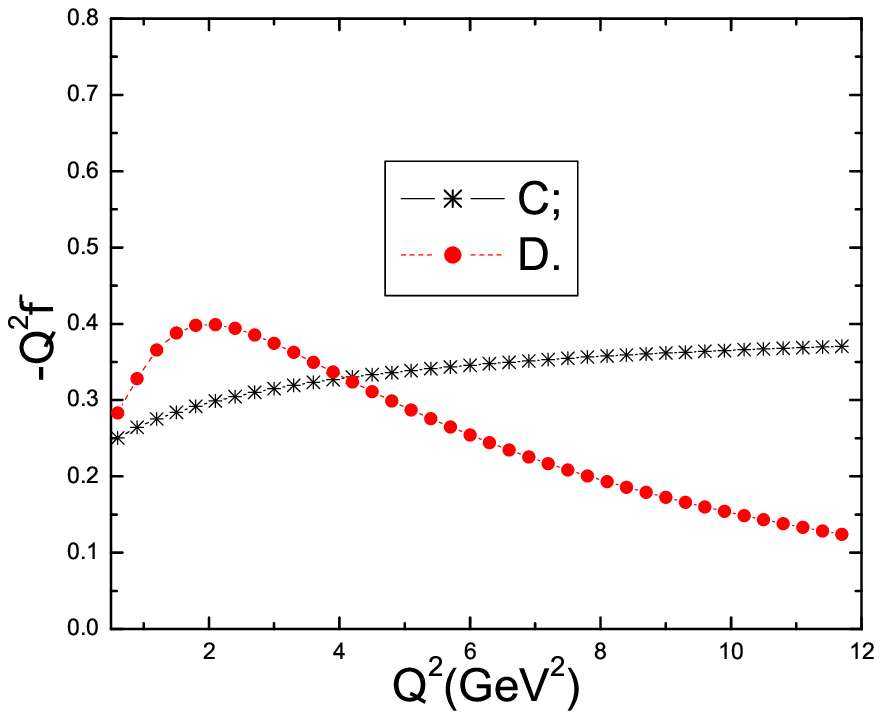}
   \caption{The central values of the $Q^2f^+_{K\pi}(Q^2)$ and $Q^2f^-_{K\pi}(Q^2)$ at $Q^2=(0-12)GeV^2$,
 A from Eq.(13), B from Eq.(14), C from Eq.(15) and Eq.(13) , D from Eq.(16) and Eq.(14). }
\end{figure}

The numerical values $|f^-_{K\pi}(0)|=5.15$ from Eq.(15) and
$|f^-_{K\pi}(0)|=11.47$ from Eq.(16) are very large comparing with
the values from the experimental data \cite{ExpKP}, ChPT
\cite{ChPTKP}, lattice QCD \cite{LattKP} and the Bethe-Salpeter
equation \cite{BSEKP}, they can not give any reliable predictions.
It is not un-expected, from the sum  rules in Eqs.(15-16), we can
see the terms  like
\begin{eqnarray}
\int_{\Delta_A}^1du
\frac{\phi_p(u)}{u}e^{-\frac{m_s^2+u(1-u)m_\pi^2-(1-u)q^2}{uM^2}} \,
,
\nonumber \\
\int_{\Delta_B}^1du
\frac{\phi_p(u)}{u}e^{-\frac{m_d^2+u(1-u)m_K^2-(1-u)q^2}{uM^2}}\,
.\nonumber
\end{eqnarray}
The $f^-_{K\pi}(Q^2)$ are greatly enhanced in the region of small
$Q^2$ due to the extra $\frac{1}{u}$ comparing with the
corresponding $f^+_{K\pi}(Q^2)$ in Eqs.(13-14), in the limit
$Q^2=0$, $\Delta_A\approx 0.017$ and $\Delta_B\approx0.00004$, the
dominant contributions come from the end-point of the light-cone
distribution amplitudes, such a infrared behavior can spoil the
extrapolation; we should introduce extra phenomenological
form-factors (for example, the Sudakov factor \cite{HnLi}) to
suppress the contribution from the end-point, that may be our next
work.
 Our numerical values of $f^-_{K\pi}(Q^2)$
have a negative sign to the ones of the $f^+_{K\pi}(Q^2)$, which is
consistent with the existing theoretical calculations and
experimental data.  In Figs.6-7, we plot the form factor
$f^-_{K\pi}(Q^2)$ at the momentum range $Q^2=(0-12)GeV^2$, from the
figures, we can see that the form factor $Q^2f^-_{K\pi}(Q^2)$ from
Eq.(15), just like the $Q^2f^+_{K\pi}(Q^2)$ from Eq.(13), is rather
flat at $Q^2 > 5 GeV^2 $, which means that at large momentum
transfers, the $f^-_{K\pi}(Q^2)\sim \frac{1}{Q^2}$, it is also
expected from the naive power counting rules \cite{Brodsky},  the
terms proportional to $\frac{1}{Q^{2n}}$ with $n\geq 2$ canceled out
with each other. At  momentum transfers with $Q^2>5GeV^2$, the terms
of the $f^-_{K\pi} (Q^2)$ from Eq.(16) proportional to
$\frac{1}{Q^{2n}}$ with $n\geq 2$ manifest themselves, which results
in the values of the $Q^2f^-_{K\pi}(Q^2)$ (just like the
$Q^2f^+_{K\pi}(Q^2)$ from Eq.(14)) decrease with the increase of the
$Q^2$.

 In the light-cone QCD sum rules, we carry out the
operator product expansion near the light-cone $x^2\approx 0$, which
corresponds to the $Q^2\gg 0$ and $P^2 \gg 0 $. The four sum rules
in Eqs.(13-16) can be taken as   some functions which model the
vector form factors $ f^+_{K\pi}(Q^2) $ and $f^-_{K\pi} (Q^2)$  at
large momentum transfers,    we extrapolate the $ f^+_{K\pi}(Q^2) $
and $f^-_{K\pi} (Q^2)$  to the zero momentum transfer  or beyond
with the analytical continuation\footnote{We can borrow some ideas
from the electromagnetic form-factor of the $\pi$-photon
$f_{\gamma^* \pi^0 }(Q^2)$, the  value of the $f_{\gamma^* \pi^0
}(0)$ is fixed by the partial conservation of the axial current and
the effective  anomaly lagrangian, $f_{\gamma^* \pi^0 }(0) =
\frac1{\pi f_{\pi}}$, in the limit large-$Q^2$, the perturbative QCD
predicts that $f_{\gamma^* \pi^0 }(Q^2) =4\pi f_{\pi}/Q^2 $.
 The Brodsky-Lepage interpolation formula \cite{BJ81}
 \begin{eqnarray}
f_{\gamma^* \pi^0 }(Q^2) = \frac{1}{ \pi f_{\pi} \left [1+Q^2/(4
\pi^2 f_{\pi}^2) \right ]} =\frac{1}{ \pi f_\pi (1+Q^2/s_0) }
\nonumber
\end{eqnarray}
 can reproduce both the value of $Q^2 =0$
  and the  behavior of large-$Q^2$, the energy scale $s_0$ ($s_0 = 4 \pi^2
f_{\pi}^2 \approx 0.67 \, GeV^2$) is numerically   close to   the
squared mass of the $\rho$ meson, $m_{\rho}^2 \approx 0.6 \, GeV^2$.
The Brodsky-Lepage interpolation formula is similar to the result of
the vector meson dominance, $f_{\gamma^* \pi^0 }(Q^2) = 1/\left\{\pi
f_\pi (1+Q^2/m_{\rho}^2)\right\}$. In the vector meson dominance
approach, the calculation is performed at the time-like energy scale
$q^2<1GeV^2$ and the electromagnetic  current is saturated by the
vector meson $\rho$, where the mass $m_{\rho}$ serves as a parameter
determining the pion charge radius. With a slight modification of
the mass parameter, $m_\rho=\Lambda_\pi=776MeV$, the experimental
data can be well described by the single-pole formula  at the
interval $Q^2=(0-10)GeV^2$ \cite{CLEO97}. In Ref.\cite{WangSN07},
the four form-factors of the $\Sigma \to n$ have satisfactory
behaviors at large $Q^2$ which are expected by the naive power
counting rules, and have finite values at $Q^2=0$, the analytical
expressions of the four form factors $f_1(Q^2)$, $f_2(Q^2)$,
$g_1(Q^2)$ and $g_2(Q^2)$ are taken as some Brodsky-Lepage  type
interpolation formulaes, although they are calculated at rather
large $Q^2$, the extrapolation to the lower energy transfers has no
solid theoretical foundation. The numerical values of  the
   $f_1(0)$, $f_2(0)$, $g_1(0)$ and $g_2(0)$  are
 compatible with the experimental data and  theoretical
 calculations (in magnitude). In this article, the vector form factors $f_{K\pi}^+(Q^2)$ and  $f_{K\pi}^-(Q^2)$ can also
 be taken as some Brodsky-Lepage  type interpolation formulaes, the low momentum transfer $Q^2$ behaviors
  may be good or bad.}.
The chosen functions may have good or bad lower $Q^2$ behaviors,
which correspond to the systematic errors, more experimental data
are needed to select the ideal ones.

\section{Conclusions}

In this article, we calculate the vector form factors
$f^+_{K\pi}(Q^2)$ and $f^-_{K\pi}(Q^2)$ within the framework of the
light-cone QCD sum rules approach. The $f^+_{K\pi}(0)$ is the basic
input parameter in extracting the CKM matrix element $|V_{us}|$ from
the $K_{\ell 3}$ decays. The numerical values of the
$f^+_{K\pi}(Q^2)$ are compatible with the existing theoretical
calculations, the central value $f^+_{K\pi}(0)=0.97$ is in excellent
agreement with the values from the ChPT and lattice QCD. The values
of the $|f^-_{K\pi}(0)|$ are very large comparing with the
theoretical calculations and experimental data, and can not give any
reliable predictions. At large momentum transfers with $Q^2>
5GeV^2$, the form factors $f^+_{K\pi}(Q^2)$ and $|f^-_{K\pi}(Q^2)|$
can either take up the asymptotic behavior of $\frac{1}{Q^2}$ or
decrease more quickly than $\frac{1}{Q^2}$, more experimental data
are needed to select the ideal sum rules.

\section*{Appendix}

The light-cone distribution amplitudes of the $K$ meson,
\begin{eqnarray}
\langle0| {\bar u} (0) \gamma_\mu \gamma_5 s(x) |K(p)\rangle& =& i
f_K p_\mu \int_0^1 du  e^{-i u p\cdot x}
\left\{\varphi_K(u)+\frac{m_K^2x^2}{16}
A(u)\right\}\nonumber\\
&&+if_K m_K^2\frac{x_\mu}{2p\cdot x}
\int_0^1 du  e^{-i u p \cdot x} B(u) \, , \nonumber\\
\langle0| {\bar u} (0) i \gamma_5 s(x) |K(p)\rangle &=& \frac{f_K
m_K^2}{ m_s+m_u}
\int_0^1 du  e^{-i u p \cdot x} \varphi_p(u)  \, ,  \nonumber\\
\langle0| {\bar u} (0) \sigma_{\mu \nu} \gamma_5 s(x) |K(p)\rangle
&=&i(p_\mu x_\nu-p_\nu x_\mu)  \frac{f_K m_K^2}{6 (m_s+m_u)}
\int_0^1 du
e^{-i u p \cdot x} \varphi_\sigma(u) \, ,  \nonumber\\
\langle0| {\bar u} (0) \sigma_{\alpha \beta} \gamma_5 g_s G_{\mu
\nu}(v x)s(x) |K(p)\rangle&=& f_{3 K}\left\{(p_\mu p_\alpha
g^\bot_{\nu
\beta}-p_\nu p_\alpha g^\bot_{\mu \beta}) -(p_\mu p_\beta g^\bot_{\nu \alpha}\right.\nonumber\\
&&\left.-p_\nu p_\beta g^\bot_{\mu \alpha})\right\} \int {\cal
D}\alpha_i \varphi_{3 K} (\alpha_i)
e^{-ip \cdot x(\alpha_s+v \alpha_g)} \, ,\nonumber\\
\langle0| {\bar u} (0) \gamma_{\mu} \gamma_5 g_s G_{\alpha
\beta}(vx)s(x) |K(p)\rangle&=&  p_\mu  \frac{p_\alpha
x_\beta-p_\beta x_\alpha}{p
\cdot x}f_Km_K^2\nonumber\\
&&\int{\cal D}\alpha_i A_{\parallel}(\alpha_i) e^{-ip\cdot
x(\alpha_s +v \alpha_g)}\nonumber \\
&&+ f_Km_K^2 (p_\beta g_{\alpha\mu}-p_\alpha
g_{\beta\mu})\nonumber\\
&&\int{\cal D}\alpha_i A_{\perp}(\alpha_i)
e^{-ip\cdot x(\alpha_s +v \alpha_g)} \, ,  \nonumber\\
\langle0| {\bar u} (0) \gamma_{\mu}  g_s \tilde G_{\alpha
\beta}(vx)s(x) |K(p)\rangle&=& p_\mu  \frac{p_\alpha x_\beta-p_\beta
x_\alpha}{p \cdot
x}f_Km_K^2\nonumber\\
&&\int{\cal D}\alpha_i V_{\parallel}(\alpha_i) e^{-ip\cdot
x(\alpha_s +v \alpha_g)}\nonumber \\
&&+ f_Km_K^2 (p_\beta g_{\alpha\mu}-p_\alpha
g_{\beta\mu})\nonumber\\
&&\int{\cal D}\alpha_i V_{\perp}(\alpha_i) e^{-ip\cdot x(\alpha_s +v
\alpha_g)} \, ,
\end{eqnarray}
here the operator $\tilde G_{\alpha \beta}$  is the dual of the
$G_{\alpha \beta}$, $\tilde G_{\alpha \beta}= {1\over 2}
\epsilon_{\alpha \beta  \mu\nu} G^{\mu\nu} $, ${\cal{D}}\alpha_i$ is
defined as ${\cal{D}} \alpha_i =d \alpha_1 d \alpha_2 d \alpha_3
\delta(1-\alpha_1 -\alpha_2 -\alpha_3)$. The  light-cone
distribution amplitudes are parameterized as
\begin{eqnarray}
\phi_K(u,\mu)&=&6u(1-u)
\left\{1+a_1C^{\frac{3}{2}}_1(2u-1)+a_2C^{\frac{3}{2}}_2(2u-1)
+a_4C^{\frac{3}{2}}_4(2u-1)\right\}\, , \nonumber\\
\varphi_p(u,\mu)&=&1+\left\{30\eta_3-\frac{5}{2}\rho^2\right\}C_2^{\frac{1}{2}}(2u-1)\nonumber \\
&&+\left\{-3\eta_3\omega_3-\frac{27}{20}\rho^2-\frac{81}{10}\rho^2 a_2\right\}C_4^{\frac{1}{2}}(2u-1)\, ,  \nonumber \\
\varphi_\sigma(u,\mu)&=&6u(1-u)\left\{1
+\left[5\eta_3-\frac{1}{2}\eta_3\omega_3-\frac{7}{20}\rho^2-\frac{3}{5}\rho^2 a_2\right]C_2^{\frac{3}{2}}(2u-1)\right\}\, , \nonumber \\
T(\alpha_i,\mu) &=& 360 \alpha_u \alpha_s \alpha_g^2 \left \{1
+\lambda_3(\alpha_u-\alpha_s)+ \omega_3 \frac{1}{2} ( 7 \alpha_g
- 3) \right\} \, , \nonumber\\
V_{\parallel}(\alpha_i,\mu) &=& 120\alpha_u \alpha_s \alpha_g \left(
v_{00}+v_{10}(3\alpha_g-1)\right)\, ,
\nonumber \\
A_{\parallel}(\alpha_i,\mu) &=& 120 \alpha_u \alpha_s \alpha_g
a_{10} (\alpha_s-\alpha_u)\, ,
\nonumber\\
V_{\perp}(\alpha_i,\mu) &=& -30\alpha_g^2
\left\{h_{00}(1-\alpha_g)+h_{01}\left[\alpha_g(1-\alpha_g)-6\alpha_u
\alpha_s\right] \right.  \nonumber\\
&&\left. +h_{10}\left[
\alpha_g(1-\alpha_g)-\frac{3}{2}\left(\alpha_u^2+\alpha_s^2\right)\right]\right\}\,
, \nonumber\\
A_{\perp}(\alpha_i,\mu) &=&  30 \alpha_g^2 (\alpha_u-\alpha_s) \left\{h_{00}+h_{01}\alpha_g+\frac{1}{2}h_{10}(5\alpha_g-3)  \right\}, \nonumber\\
A(u,\mu)&=&6u(1-u)\left\{
\frac{16}{15}+\frac{24}{35}a_2+20\eta_3+\frac{20}{9}\eta_4 \right.
\nonumber \\
&&+\left[
-\frac{1}{15}+\frac{1}{16}-\frac{7}{27}\eta_3\omega_3-\frac{10}{27}\eta_4\right]C^{\frac{3}{2}}_2(2u-1)
\nonumber\\
&&\left.+\left[
-\frac{11}{210}a_2-\frac{4}{135}\eta_3\omega_3\right]C^{\frac{3}{2}}_4(2u-1)\right\}+\left\{
 -\frac{18}{5}a_2+21\eta_4\omega_4\right\} \nonumber\\
 && \left\{2u^3(10-15u+6u^2) \log u+2\bar{u}^3(10-15\bar{u}+6\bar{u}^2) \log \bar{u}
 \right. \nonumber\\
 &&\left. +u\bar{u}(2+13u\bar{u})\right\} \, ,\nonumber\\
 g_K(u,\mu)&=&1+g_2C^{\frac{1}{2}}_2(2u-1)+g_4C^{\frac{1}{2}}_4(2u-1)\, ,\nonumber\\
 B(u,\mu)&=&g_K(u,\mu)-\phi_K(u,\mu)\, ,
\end{eqnarray}
where
\begin{eqnarray}
h_{00}&=&v_{00}=-\frac{\eta_4}{3} \, ,\nonumber\\
a_{10}&=&\frac{21}{8}\eta_4 \omega_4-\frac{9}{20}a_2 \, ,\nonumber\\
v_{10}&=&\frac{21}{8}\eta_4 \omega_4 \, ,\nonumber\\
h_{01}&=&\frac{7}{4}\eta_4\omega_4-\frac{3}{20}a_2 \, ,\nonumber\\
h_{10}&=&\frac{7}{2}\eta_4\omega_4+\frac{3}{20}a_2 \, ,\nonumber\\
g_2&=&1+\frac{18}{7}a_2+60\eta_3+\frac{20}{3}\eta_4 \, ,\nonumber\\
g_4&=&-\frac{9}{28}a_2-6\eta_3\omega_3 \, ,
\end{eqnarray}
 here  $ C_2^{\frac{1}{2}}$, $ C_4^{\frac{1}{2}}$
 and $ C_2^{\frac{3}{2}}$ are Gegenbauer polynomials,
  $\eta_3=\frac{f_{3K}}{f_K}\frac{m_q+m_s}{M_K^2}$ and  $\rho^2={m_s^2\over M_K^2}$
 \cite{LCSR,LCSRreview,Belyaev94,Ball98,Ball06}.

\section*{Acknowledgments}
This  work is supported by National Natural Science Foundation,
Grant Number 10405009,  and Key Program Foundation of NCEPU.

\end{document}